\documentclass[12pt]{article}

\usepackage{graphicx}
\usepackage{a4}

\usepackage{fancybox}
\usepackage{epsfig}
\usepackage{color}

\usepackage{amsfonts}
\usepackage{mathrsfs}
\usepackage{amssymb}
\usepackage{amsmath}

\usepackage{dcolumn}
\usepackage{bm}

\def\pa{\partial}

\def\al{\alpha}
\def\be{\beta}

\def\ep{\epsilon}

\def\th{\theta}

\def\si{\sigma}

\def\Om{\Omega}

\newcommand{\ben}{\begin{equation}}
\newcommand{\een}{\end{equation}}
\newcommand{\bea}{\begin{eqnarray}}
\newcommand{\eea}{\end{eqnarray}}
\newcommand{\ba}{\begin{array}}
\newcommand{\ea}{\end{array}}
\newcommand{\bit}{\begin{itemize}}
\newcommand{\eit}{\end{itemize}}

\newcommand{\vs}[1]{\vspace{#1 mm}}

\newcommand{\dsl}{\pa \kern-0.5em /}

\begin{document}

\topmargin 0pt \oddsidemargin 0mm

\vspace{2mm}

\begin{center}

{\Large \bf The force-free dipole magnetosphere in non-linear
electrodynamics}

\vs{10}

 {\large Huiquan Li \footnote{E-mail: lhq@ynao.ac.cn}, Xiaolin Yang and Jiancheng Wang}

\vspace{6mm}

{\em

Yunnan Observatories, Chinese Academy of Sciences, \\
650216 Kunming, China

Key Laboratory for the Structure and Evolution of Celestial Objects,
\\ Chinese Academy of Sciences, 650216 Kunming, China

Center for Astronomical Mega-Science, Chinese Academy of Sciences,
\\ 100012 Beijing, China}

\end{center}

\vs{9}

\begin{abstract}

Quantum electrodynamics (QED) effects may be included in physical
processes of magnetar and pulsar magnetospheres with strong magnetic
fields. Involving the quantum corrections, the Maxwell
electrodynamics is modified to non-linear electrodynamics. In this
work, we study the force-free magnetosphere in non-linear
electrodynamics in a general framework. The pulsar equation
describing a steady and axisymmetric magnetosphere is derived, which
now admits solutions with corrections. We derive the first-order
non-linear corrections to the near-zone dipole magnetosphere in some
popular non-linear effective theories. The field lines of the
corrected dipole tend to converge on the rotational axis so that the
fields in the polar region are stronger compared to the pure dipole
case.

\end{abstract}

%\textit{Keywords: pulsar magnetosphere, magnetic fields, neutron stars}

%\textit{PACS:}

\section{Introduction}
\label{sec:introduction}
%%%%%%%%%%%%%%%%%%%%%%%%%%%%%%%%%%%%%%%%%%%%%%%%%%%%%%%%%%%%%%%%%%%%%%%%%%%%

It is inferred that pulsars and magnetars possess very strong
magnetic fields. The field strength can even exceed the critical
value $B_Q=m_e^2c^3/\hbar e\simeq 4.4\times10^{13}$ G, above which
the quantum electrodynamics (QED) effects should not be ignored.

QED effects will affect the polarization and spectra of the thermal
radiation from the surface of magnetars and pulsars with strong
magnetic fields. In the magnetospheres, the photon polarizations can
be decomposed into two modes according to the propagating direction
and the local magnetic field direction. Due to the resonance between
the vacuum and plasma birefringence, one of the polarization modes
of the X-ray photon can be converted into the other. This gives rise
to an energy-dependent polarization signature for the observed
quiescent non-thermal X-ray emission
\cite{Lai:2003nd,Denisov:2003ba,Harding:2006qn,vanAdelsberg:2006uu,
Fernandez:2011aa,Kaspi:2017fwg,Krawczynski:2019ofl}. The magnetar
magnetospheres are opaque to high energy photons due to the
attenuation by the magnetic photon splitting (below the energy
threshold $2m_ec^2$) and pair production (above the threshold). This
will distort the blackbody spectra of the surface thermal radiation
and can be tested with future precise observations of the spectra
and the polarizations \cite{Wadiasingh:2019jdr,Hu:2019nyw}, which
meanwhile provides information on the surface magnetic fields in the
magnetar magnetospheres.

Thus, the strength and geometry of the magnetic fields are crucial
in these QED processes. It is expected that these processes more
effectively work at low altitudes where the magnetic fields are
higher. In the near-zone regions, the geometry of the pulsar
magnetospheres is usually taken to be of a force-free dipole
structure. When including the QED corrections, the Maxwell
electrodynamics should be modified with additional non-linear terms
and the dipole magnetosphere must be corrected with non-linear
contributions, as analyzed in previous works
\cite{Heyl:1997hs,Petri:2015ekv}.

In this work, we consider the force-free magnetosphere in general
non-linear electrodynamics. In contrast to previous treatments
\cite{Freytsis:2015qda,Petri:2016gac}, we shall follow the
traditional approach to do so. We first derive the pulsar equation,
describing the steady and axisymmetric magnetospheres, in non-linear
electrodynamics. We then obtain the corrected dipole magnetosphere
to leading order from the pulsar equation in some popular non-linear
effective theories, like the Euler-Heisenberg (EH) theory,
Born-Infeld (BI) theory and the logarithmic theory.

%%%%%%%%%%%%%%%%%%%%%%%%%%%%%%%%%%%%%%%%%%%%%%%%%%%%%%%%%%%%%%%%%%%%%%%%%%%%
\section{Non-linear electrodynamics}
\label{sec:NLED}
%%%%%%%%%%%%%%%%%%%%%%%%%%%%%%%%%%%%%%%%%%%%%%%%%%%%%%%%%%%%%%%%%%%%%%%%%%%%

The action of general electrodynamics takes the form
\begin{equation}
 S=\int \sqrt{-g}\left[\frac{1}{4\pi}\mathcal{L}_{\textrm{EM}}(s,p)
+A_\mu J^\mu\right]d^4x,
\end{equation}
where $\mathcal{L}_{\textrm{EM}}(s,p)$ is the general Lagrangian of
the electromagnetic fields with
\begin{equation}
 s\equiv\frac{1}{4}F^{\mu\nu}F_{\mu\nu}
=\frac{1}{2}(\mathbf{B}^2-\mathbf{E}^2),
\end{equation}
\begin{equation}
 p\equiv\frac{1}{4}\widetilde{F}^{\mu\nu}F_{\mu\nu}=\mathbf{E}\cdot
\mathbf{B}.
\end{equation}
The dual field strength
$\widetilde{F}^{\mu\nu}=(1/2)\ep^{\mu\nu\rho\si}F_{\rho\si}$.

The equations of motion can be derived from the action. It is
obvious that the Bianchi identity is automatically satisfied:
\begin{equation}
 \nabla_\mu \widetilde{F}^{\mu\nu}=0.
\end{equation}
In Minkowski spacetime, the equation can be decomposed into
\begin{equation}
 \nabla\times\mathbf{E}=-\dot{\mathbf{B}},
\end{equation}
\begin{equation}
 \nabla\cdot\mathbf{B}=0.
\end{equation}
The relation between the current and fields is given by
\begin{equation}\label{e:currents}
 \nabla_\mu G^{\mu\nu}=4\pi J^\nu,
\end{equation}
where $J^\nu$ is the conserved current and
\begin{equation}
 G^{\mu\nu}=SF^{\mu\nu}+P\widetilde{F}^{\mu\nu},
\end{equation}
with
\begin{equation}
 S\equiv\pa_s\mathcal{L}_{\textrm{EM}}, \textrm{ }
\textrm{ }\textrm{ } P\equiv\pa_p\mathcal{L}_{\textrm{EM}}.
\end{equation}
When $S=-1$ and $P=0$, the equation reduces to the Maxwell theory
case. In Minkowski spacetime, the equation can be re-expressed as:
\begin{equation}
 \nabla\cdot\mathbf{D}=4\pi\rho,
\end{equation}
\begin{equation}
 \nabla\times\mathbf{H}=4\pi\mathbf{j}+\dot{\mathbf{D}},
\end{equation}
where
\begin{equation}
 \mathbf{D}=-S\mathbf{E}+P\mathbf{B},
\end{equation}
\begin{equation}
 \mathbf{H}=-S\mathbf{B}-P\mathbf{E}.
\end{equation}

%%%%%%%%%%%%%%%%%%%%%%%%%%%%%%%%%%%%%%%%%%%%%%%%%%%%%%%%%%%%%%%%%%%%%%%%%%%%
\section{The force-free condition}
\label{sec:ff}
%%%%%%%%%%%%%%%%%%%%%%%%%%%%%%%%%%%%%%%%%%%%%%%%%%%%%%%%%%%%%%%%%%%%%%%%%%%%

The derivative of the Lagrangian with respect to the metric gives
the energy-momentum tensor of the electromagnetic fields
\begin{equation}
 T_{\textrm{EM}}^{\mu\nu}
=-\frac{1}{4\pi}[SF^\mu_{\textrm{ }\textrm{ }\al}F^{\nu\al}+
P\widetilde{F}^\mu_{\textrm{ }\textrm{
}\al}F^{\nu\al}-g^{\mu\nu}\mathcal{L}_{\textrm{EM}}].
\end{equation}
The tensor satisfies
\begin{equation}\label{e:divtensor}
 \nabla_\mu T_{\textrm{EM}}^{\mu\nu}=J_\mu F^{\mu\nu}.
\end{equation}
The equation (\ref{e:divtensor}) relating the divergence of the EM
energy-momentum to the Lorentz force takes the same form as in the
Maxwell theory. It determines the change of the momenta of the
charged particles in the system.

It is usually assumed that, in a steady magnetosphere filled with
plasma, the charged particles in the magnetospheres with the strong
EM fields should feel no net force (at least in most regions). This
means that the Lorentz force in Eq.\ (\ref{e:divtensor}) should
vanish:
\begin{equation}\label{e:ffcond}
 J_\mu F^{\mu\nu}=0.
\end{equation}
This is the force-free condition in general non-linear
electrodynamics, also of the same form as in the Maxwell theory.
This condition also says that the dynamics of the energy density in
the system is dominated by the electromagnetic fields and the
inertial of the plasma in the system can be ignored. That is,
$T_{\textrm{EM}}^{\mu\nu}$ can be approximately taken as the
energy-momentum density of the whole system so that it is conserved.

In components, the force-free equation is decomposed into
\begin{equation}
 \mathbf{j}\cdot\mathbf{E}=0,
\textrm{ }\textrm{ }\textrm{ } \rho
\mathbf{E}+\mathbf{j}\times\mathbf{B}=0,
\end{equation}
which implies
\begin{equation}
 p=\mathbf{E}\cdot\mathbf{B}=0.
\end{equation}
So we simply have
$\mathcal{L}_{\textrm{EM}}(s,p)=\mathcal{L}_{\textrm{EM}}(s)$ and
$G^{\mu\nu}=SF^{\mu\nu}$ under the force-free condition.

%%%%%%%%%%%%%%%%%%%%%%%%%%%%%%%%%%%%%%%%%%%%%%%%%%%%%%%%%%%%%%%%%%%%%%%%%%%%
\section{The pulsar equation}
\label{sec:PE}
%%%%%%%%%%%%%%%%%%%%%%%%%%%%%%%%%%%%%%%%%%%%%%%%%%%%%%%%%%%%%%%%%%%%%%%%%%%%

Under the force-free condition, the equations describing a
force-free magnetosphere can be derived. As usual, we consider the
simplest case: the axisymmetric and steady magnetospheres in
Minkowski spacetime. On the spherical coordinates, the force-free
condition (\ref{e:ffcond}) reads:
\begin{equation}\label{e:ffcond1}
 \pa_r A_0J^r+\pa_\th A_0 J^\th=0,
\end{equation}
\begin{equation}\label{e:ffcond2}
 \pa_r A_0J^0+F_{r\th}J^\th+\pa_r A_\phi J^\phi=0,
\end{equation}
\begin{equation}\label{e:ffcond3}
 \pa_\th A_0J^0-F_{r\th}J^r+\pa_\th A_\phi J^\phi=0,
\end{equation}
\begin{equation}\label{e:ffcond4}
 \pa_r A_\phi J^r+\pa_\th A_\phi J^\th=0.
\end{equation}

For convenience, let us define the Poison bracket as in
\cite{Petrova:2016whc} (and also \cite{Compere:2016xwa}):
\begin{equation}
 [A, B]\equiv\mathcal{L}_T B=\pa_r A\pa_\th B-\pa_\th A\pa_r B,
\end{equation}
where the tangent vector $T=\pa_r A\pa_\th-\pa_\th A\pa_r$. The
necessary condition that $A$ is a function of $B$ (vice versa) is
that $[A,B]=0$.

From Eqs.\ (\ref{e:ffcond1}) and (\ref{e:ffcond4}), we can find that
\begin{equation}
 [A_0, A_\phi]=0.
\end{equation}
So $A_0$ should be a function of $A_\phi$. We can define:
\begin{equation}
 dA_0=-\Om(A_\phi)dA_\phi.
\end{equation}
As known, $\Om$ is the angular velocity of a magnetic field line. It
is constant along the magnetic field line.

The Maxwell equations (\ref{e:currents}) with non-linear corrections
are expressed as
\begin{equation}\label{e:currents1}
 J^0=-\frac{1}{4\pi}\nabla\cdot\left(S\nabla A_0\right),
\end{equation}
\begin{equation}\label{e:currents2}
 J^r=-\frac{1}{4\pi r^2\sin\th}\pa_\th(\sin\th SF_{r\th}),
\end{equation}
\begin{equation}\label{e:currents3}
 J^\th=\frac{1}{4\pi r^2\sin\th}\pa_r(\sin\th SF_{r\th}),
\end{equation}
\begin{equation}\label{e:currents4}
 J^\phi=\frac{1}{4\pi}\nabla\cdot\left(\frac{S\nabla A_\phi}
{r^2\sin^2\th}\right).
\end{equation}

From Eqs.\ (\ref{e:ffcond1}), (\ref{e:ffcond4}), (\ref{e:currents2})
and (\ref{e:currents3}), we find that
\begin{equation}
 [A_0,\textrm{}\sin\th SF_{r\th}]=[A_\phi,\textrm{}\sin\th SF_{r\th}]=0.
\end{equation}
So $\sin\th SF_{r\th}$ is also a function of $A_\phi$. Let us denote
\begin{equation}
 \psi\equiv2\pi A_\phi, \textrm{ }\textrm{ }\textrm{ }
I(\psi)\equiv-2\pi\sin\th SF_{r\th}.
\end{equation}
Then the charge density is expressed as
\begin{equation}
 J^0=\frac{1}{8\pi^2}\nabla\cdot\left(S\Om\nabla\psi\right).
\end{equation}
From Eq.\ (\ref{e:ffcond2}) or (\ref{e:ffcond3}), we have
\begin{equation}\label{e:jphi}
 J^\phi=\Om J^0-\frac{II'}{8\pi^2r^2\sin^2\th S},
\end{equation}
where the prime denotes the derivative with respect to $\psi$.

By comparing Eqs.\ (\ref{e:currents4}) and (\ref{e:jphi}), we can
derive the general pulsar equation:
\begin{equation}\label{e:pulsareq}
 S\nabla\cdot\left(\frac{S\nabla\psi}{r^2\sin^2\th}\right)
-S\Om\nabla\cdot\left(S\Om\nabla\psi\right)
=-\frac{II'}{r^2\sin^2\th}.
\end{equation}
Specifically, the equation on spherical coordinates can be written
as
\begin{eqnarray}\label{e:pulsareqsph}
 \frac{1}{r^2}(1-r^2\sin^2\th\Om^2)\left[r^2S\pa_r(S\pa_r\psi)
+S\pa_\th(S\pa_\th\psi)\right]-\sin^2\th\Om\Om'S^2
[r^2(\pa_r\psi)^2+(\pa_\th\psi)^2]
\nonumber \\
-2r\sin^2\th\Om^2S^2\pa_r\psi-\frac{1}{r^2}(1+r^2\sin^2\th\Om^2)
\cot\th S^2\pa_\th\psi=-II'.
\end{eqnarray}
When $S=-1$, this reduces to the pulsar equation in Maxwell's
theory.

The electromagnetic fields in the unit basis of spherical
coordinates are expressed as:
\begin{equation}
 \mathbf{D}=-S\mathbf{E}=\frac{S\Om}{2\pi r}\left(
r\pa_r\psi,\pa_\th\psi,0\right),
\end{equation}
\begin{equation}
 \mathbf{H}=-S\mathbf{B}=\frac{1}{2\pi r^2\sin\th}\left(
-S\pa_\th\psi,rS\pa_r\psi,rI\right).
\end{equation}
With them, we have
\begin{eqnarray}
 s=\frac{1}{8\pi^2r^4\sin^2\th}\left\{\frac{r^2I^2}{S^2}+
(1-r^2\sin^2\th\Om^2)[r^2(\pa_r\psi)^2 +(\pa_\th\psi)^2]\right\}
\nonumber \\
=\frac{1}{8\pi^2x^2}\left\{\frac{I^2}{S^2}+
(1-x^2\Om^2)[(\pa_x\psi)^2 +(\pa_z\psi)^2]\right\},
\end{eqnarray}
which must be non-negative. The expression on the cylinder
coordinates ($x=r\sin\th$ and $z=r\cos\th$) given in the second line
indicates that the translational symmetry along the rotation axis
remains in non-linear electrodynamics, i.e., the action and the
equations are invariant with the transformation: $z\rightarrow
z'=z+\ep$.

The spin-down rate is obtained:
\begin{equation}
 L=\int \mathbf{S}\cdot d\mathbf{s}
=-\frac{1}{8\pi^2}\int I(\psi)\Om(\psi)d\psi,
\end{equation}
where the Poynting flux is
\begin{equation}
 \mathbf{S}=\frac{1}{4\pi}\mathbf{E}\times\mathbf{H}.
\end{equation}
So the torque takes the same form as in the Maxwell theory.

%%%%%%%%%%%%%%%%%%%%%%%%%%%%%%%%%%%%%%%%%%%%%%%%%%%%%%%%%%%%%%%%%%%%%%%%%%%%
\section{The near-zone dipole magnetospheres}
\label{sec:dipole}
%%%%%%%%%%%%%%%%%%%%%%%%%%%%%%%%%%%%%%%%%%%%%%%%%%%%%%%%%%%%%%%%%%%%%%%%%%%%

The pulsar equation is hard to solve even in the Maxwell theory. So
it is expected that numerical methods are needed to solve the
equation (\ref{e:pulsareq}) in non-linear electrodynamics. But here
we do not need to seek for the global solutions. We only need to
focus on the magnetosphere at low altitudes where the
electromagnetic fields are strong and the non-linear corrections may
be important.

As done in the Maxwell theory, the near-zone magnetospheres on
pulsars are usually taken as a dipole structure, which serves as the
inner boundary condition in numerical simulations of pulsar
magnetospheres
\cite{Michel:1973ga,Contopoulos:1999ga,Gruzinov:2004jc}. This
structure can be obtained from the pulsar equation at
$r\rightarrow0$, where the rotational velocities of the magnetic
field lines are much less than the speed of light and the electric
current is negligible by setting $I=0$. In this limit, the equation
(\ref{e:pulsareqsph}) with $S=-1$ reduces to:
\begin{equation}\label{e:maxnzoneeq}
 \pa_\th^2\psi-\cot\th\pa_\th\psi+r^2\pa_r^2\psi=0.
\end{equation}

The equation is solved by the general form
\begin{equation}
 \psi=\psi_{-n}(\th)r^{-n},
\end{equation}
where $\psi_{-n}(\th)$ is related to the associated Legendre
polynomials for different $n$. For $n=0$, it is a monopole, and, for
$n=1$, it is a pure dipole $\psi=\sin^2\th/r$. The rotational
effects in outer regions just deform this basic dipole geometry.

In what follows, we determine the force-free magnetospheres in the
near regions in different non-linear theories.

\subsection{The BI theory}

The BI effective theory is a well regularized non-linear theory,
leading to finite self-energy of point-like charge and absence of
birefringence. It also arises from the worldvolume action of
D-branes in string theory. Some aspects of pulsar magnetospheres in
BI effective theory were discussed previously in
\cite{Denisov:2003ba,Pereira:2018mnn}. Here, we consider the
corrected dipole geometry in the theory.

The Lagrangian of electromagnetic fields in the BI effective theory
takes the form
\begin{equation}\label{e:BI}
 \mathcal{L}_{\textrm{EM}}(s,p)=b^2\left(1-\sqrt{1+
\frac{2s}{b^2}-\frac{p^2}{b^4}}\right),
\end{equation}
where the only parameter is of dimension of mass squared: $b=M^2$.
The lower bound of $M$ is constrained to be $4\times10^{-4}$ GeV by
PVLAS \cite{DellaValle:2014xoa} and 100 GeV by ATLAS in LHC
\cite{Ellis:2017edi} (see also \cite{Pereira:2018mnn}).

From the Lagrangian, we can obtain the expression of $S$:
\begin{equation}\label{e:BIS2}
 S^2=\frac{r^2(4\pi^2b^2r^2\sin^2\th-I^2)}{4\pi^2b^2r^4\sin^2\th
+(1-r^2\sin^2\th\Om^2)[r^2(\pa_r\psi)^2+(\pa_\th\psi)^2]}.
\end{equation}
At large distance $r$, $S^2\rightarrow1$ and so the pulsar equation
recovers the one in Maxwell's theory. Inserting it into the
non-linear pulsar equation (\ref{e:pulsareqsph}), we can basically
derive solutions. But it is difficult to do so. There even does not
exist solutions that are only dependent on $\th$ (like Michel's
monopole solution in the Maxwell theory).

Let us take the near-zone limit with approximately vanishing $\Om$
and $I$, for which the equation (\ref{e:pulsareqsph}) with
(\ref{e:BIS2}) is simplified to:
\begin{eqnarray}\label{e:BInzoneeq}
 \pa_\th^2\psi-\cot\th\pa_\th\psi+r^2\pa_r^2\psi+
\frac{1}{4\pi^2b^2r^3\sin^2\th}\left[r(\pa_r\psi)^2\pa_\th^2\psi
+r(\pa_\th\psi)^2\pa_r^2\psi\right.
\nonumber \\
\left.+r^2(\pa_r\psi)^3-2r\pa_r\psi\pa_\th\psi\pa_r\pa_\th
\psi+2\pa_r\psi(\pa_\th\psi)^2\right]=0.
\end{eqnarray}
There exist exact solutions that are independent of the parameter
$b$: $\psi=\cos\th$ ($n=0$), $\psi=r\cos\th$ ($n=1$) and
$\psi=r^2\sin^2\th$ ($n=2$), which are also solutions to the pulsar
equation in the Maxwell theory. So the non-linear terms do not alter
the non-rotating monopole solution.

\begin{figure}
    \includegraphics[width=0.33\columnwidth]{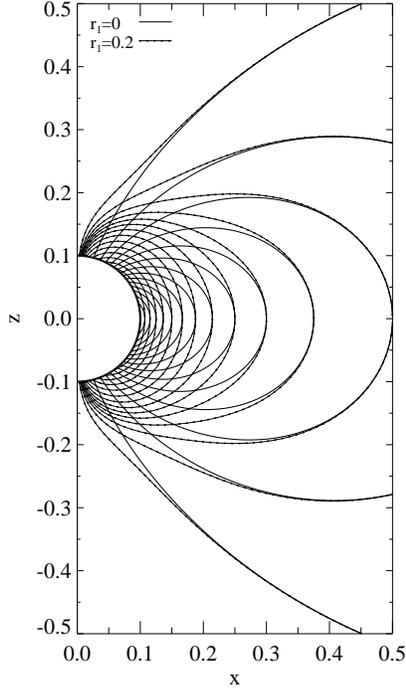}
    \caption{Magnetic field lines based on the solution (\ref{e:psi1sol})
    with different $r_1$. The radius of star is $r_*=0.1$.}
    \label{fig:f1}
\end{figure}

For the dipole solution, the situation is different. From the above
equation, we can determine that the solution can be expanded in
powers of $b^{-2}$:
\begin{equation}
 \psi=\psi^{(0)}+\psi^{(1)}+\psi^{(2)}+\cdots.
\end{equation}
The zero-th order part $\psi^{(0)}=m\sin^2\th/r$ is the pure dipole
solution. $\psi^{(1)}$ is the first order correction at the order
$\mathcal{O}(b^{-2})$. Inserting the expression into Eq.\
(\ref{e:BInzoneeq}), we get the leading order equation:
\begin{equation}\label{e:BIpsi1eq}
 \pa_\th^2\psi^{(1)}-\cot\th\pa_\th\psi^{(1)}+r^2\pa_r^2
\psi^{(1)}=\frac{3m^3\sin^2\th(1+\cos^2\th)}{4\pi^2b^2r^7}.
\end{equation}
Thus, the dependence of $\psi^{(1)}$ on $r$ should be of the form
$\sim r^{-7}$. Up to the first order, the final solution is:
\begin{equation}\label{e:psi1sol}
 \psi=\frac{m\sin^2\th}{r}\left[1+\frac{r_1^6}{r^6}
\left(1-\frac{9}{16}\sin^2\th\right)+\cdots\right].
\end{equation}
where the first-order characteristic distance
\begin{equation}
 r_1=\left(\frac{m}{\sqrt{33}\pi b}\right)^{\frac{1}{3}}.
\end{equation}
So the correction becomes unimportant sharply at $r\gg r_1$. The
first-order corrected part $\psi^{(1)}$ becomes important for
$r_1\gtrsim r\gg r_2$, where $r_2$ is the characteristic distance of
$\psi^{(2)}$ (not derived here). The distribution of the magnetic
field lines from the solution is displayed in Fig.\ (\ref{fig:f1}).
Compared with the dipole magnetosphere, the field lines around
$r\sim r_1$ tends to converge on the rotation axis. The curvature of
the field lines becomes larger in the distance less than but near
the characteristic distance $r_1$.

\subsection{The Logarithmic theory}

In the Logarithmic theory (e.g., see
\cite{Gaete:2013dta,Guo:2017bru}), the self-energy of point-like
charge is finite and the birefringent phenomenon appears. Its action
takes a logarithmic form
\begin{equation}\label{e:Log}
 \mathcal{L}_{\textrm{EM}}(s,p)=-b^2\ln\left(1+
\frac{s}{b^2}-\frac{p^2}{2b^4}\right).
\end{equation}
The equation (\ref{e:pulsareqsph}) with $I=\Om=0$ reduces to
\begin{eqnarray}
 \pa_\th^2\psi-\cot\th\pa_\th\psi+r^2\pa_r^2\psi+
\frac{1}{8\pi^2b^2r^4\sin^2\th}\left[r^2(\pa_r\psi)^2(\pa_\th^2\psi
+\cot\th\pa_\th\psi+2r\pa_r\psi-r^2\pa_r^2\psi)\right.
\nonumber \\
\left.-(\pa_\th\psi)^2(\pa_\th^2\psi-\cot\th\pa_\th\psi-4r\pa_r\psi
-r^2\pa_r^2\psi)-4r^2\pa_r\psi\pa_\th\psi\pa_r\pa_\th\psi\right]=0.
\end{eqnarray}
It is interesting that the equation has the same three exact
solutions as Eq.\ (\ref{e:BInzoneeq}). The first order solution
$\psi^{(1)}$ also takes the same form as Eq.\ (\ref{e:BIpsi1eq}). So
the geometries of the field lines are the same and the solutions can
not be discriminated in the two theories up to the first order.

\subsection{The EH theory}

The various QED effects on the physical processes in magnetar
magnetospheres, including the vacuum birefringence, photon splitting
and pair production, were mostly discussed based in the EH effective
theory \cite{Euler:1935zz,Heisenberg:1935qt}. In the weak field
limit, the Lagrangian expanded to leading orders is
\begin{equation}\label{e:EH}
 \mathcal{L}_{\textrm{EM}}(s,p)=-s+\be\left(4s^2+7p^2\right)
+\cdots,
\end{equation}
where $\be=e^2/(45 hc B_K^2)$ with $B_k=m_e^2c^3/(\hbar e)$. So the
leading order terms are the same as in the BI (\ref{e:BI}) and the
Logarithmic (\ref{e:Log}) lagrangians under the force-free
condition, just with different parameters $b$ and $\be$. The
first-order corrected dipole geometry should be also the same as
given in Fig.\ (\ref{fig:f1}). So, within the characteristic
distance, the dipole structure takes a multipole-like structure,
consistent with the corrected dipole structure to leading orders in
the EH theory \cite{Heyl:1997hs,Petri:2015ekv}.

%%%%%%%%%%%%%%%%%%%%%%%%%%%%%%%%%%%%%%%%%%%%%%%%%%%%%%%%%%%%%%%%%%%%%%%%%%%%
\section{Conclusions}
\label{sec:conclusion}
%%%%%%%%%%%%%%%%%%%%%%%%%%%%%%%%%%%%%%%%%%%%%%%%%%%%%%%%%%%%%%%%%%%%%%%%%%%%

The pulsar equation in general non-linear electrodynamics is
derived. The corrected dipole solutions in some popular non-linear
effective theories are obtained and discussed. These solutions take
the same form up to the first order, which indicate that the field
lines tend to converge on the rotation axis. So the fields are
stronger in polar region and have larger curvature within the
characteristic distance than in the pure dipole magnetosphere. This
discrepancy should be taken into account in considering the quantum
effects in the radiative transfer process of the surface emission.

%\newpage
\bibliographystyle{JHEP}
\bibliography{b}
\bibliographystyle{unsrt}

\end{document}